\documentclass[prl,superscriptaddress,twocolumn]{revtex4-1}
\usepackage{epsfig}
\usepackage{graphicx}
\usepackage{dcolumn}
\usepackage{bm}
\usepackage{tabularx}
\usepackage{hyperref}
\hypersetup{citecolor=black, filecolor= black, linkcolor= black, urlcolor= black}
\begin{document}

\title{Unexpected Stoichiometries in Mg-O System under High Pressure}

\author{Qiang Zhu}
\email{qiang.zhu@stonybrook.edu}
\affiliation{Department of Geosciences, Department of Physics and Astronomy, Stony Brook University, Stony Brook, New York 11794, USA}
\author{Artem R. Oganov}
\affiliation{Department of Geosciences, Department of Physics and Astronomy, Stony Brook University, Stony Brook, New York 11794, USA}
\affiliation{Geology Department, Moscow State University, Moscow 119992, Russia}
\author{Andriy O. Lyakhov}
\affiliation{Department of Geosciences, Department of Physics and Astronomy, Stony Brook University, Stony Brook, New York 11794, USA}

\begin{abstract}
Using \textit{ab initio} evolutionary simulations, we explore all the possible stoichiometries for Mg-O system at pressures up to 850 GPa. In addition to MgO, our calculations find that two extraordinary compounds MgO$_2$ and Mg$_3$O$_2$ become thermodynamically stable at 116 GPa and 500 GPa, respectively. Detailed chemical bonding analysis shows large charge transfer in all magnesium oxides. MgO$_2$ contains peroxide ions [O-O]$^{2-}$, while non-nuclear electron density maxima play the role of anions in the electride compound Mg$_3$O$_2$. The latter compound is calculated to have a much narrower band gap compared to MgO and MgO$_2$. We discuss conditions at which MgO$_2$ and Mg$_3$O$_2$ might exist in planetary conditions.
\end{abstract}

\keywords{density functional theory | crystal structure prediction  | high pressure | exotic compounds}

\maketitle
Magnesium oxide is one of the most abundant phases in planetary mantles, and understanding its high-pressure behavior is essential for constructing models of the Earth's and planetary interiors. For a long time, MgO was believed to be among the least polymorphic solids - only the NaCl-type structure has been observed in experiments at pressures up to 227 GPa \cite{1}. Static theoretical calculations have proposed that the NaCl-type (B1) MgO would transform into CsCl-type (B2) and the transition pressure is approximately 490 GPa at 0 K (474 GPa with the inclusion of zero-point vibration) \cite{2, 3, 4}. Calculations also predicted that MgO remains non-metallic up to extremely high pressure (20.7 TPa) \cite{3}, making it to our knowledge the most difficult mineral to metalize. Thermodynamic equilibria in the Mg-O system at 0.1 MPa pressure have been summarized in a previous study \cite{5}, concluding that only MgO is a stable composition, though metastable compounds (MgO$_2$, MgO$_4$) can be prepared at a very high oxygen fugacities.

Although MgO seems to be so simple and well understood, surprises may be in store for the researcher. For instance, it was shown recently that in the Li-H system in addition to the ``normal" LiH, new counterintuitive compounds LiH$_2$, LiH$_6$ and LiH$_8$ become stable at pressures above 100 GPa \cite{6}. If similarly unusual stoichiometries become stable also in the Mg-O system, this could have important planetological and chemical implications. In this paper, we explore this possibility and indeed find two new stoichiometries to be thermodynamically stable at high pressures. These two new stable compounds (MgO$_2$ and Mg$_3$O$_2$) exhibit interesting crystal structures with unusual chemical bonding and insulating and semiconducting electronic structures, respectively.

\section{Thermodynamically stable compounds in the Mg-O system}
We have performed structure searches with up to 20 atoms in the unit cell at pressures in the range 0-850 GPa for the Mg-O system. These searches yielded MgO as a stable oxide, but additionally two new compounds were predicted to be stable in the regions of high and low oxygen chemical potential, respectively. To confirm this and to obtain the most accurate results, we then focused search on two separate regions of chemical space; Mg-MgO and MgO-Mg, respectively. Since the structures in the two regions exhibit different properties, we report them separately.

\subsection{Exploring phase stability at high oxygen fugacities: MgO-O system and stable peroxide MgO$_2$} It is well known that monovalent (H-Cs) and divalent (Be-Ba and Zn-Hg) elements are able to form not only normal oxides, but also peroxides and even superoxides \cite{7} (for instance, BaO$_2$ has been well studied at both ambient and high pressure \cite{8, 9}). Our structure prediction calculations did confirm that magnesium peroxide exists at ambient pressure, and has a cubic pyrite-type structure, with \emph{Pa}3 symmetry and 12 atoms in the unit cell - in good agreement with experimental results \cite{10}. In this cubic phase, Mg is octahedrally coordinated by oxygen atoms (which form O$_2$ dumbbells), see Fig. 2a. However, \emph{Pa}3 MgO$_2$ ($c$-MgO$_2$ from now on) is calculated to have a positive formation enthalpy from Mg and O$_2$, and is therefore metastable. The calculation shows that on increasing pressure, $c$-MgO$_2$ transforms into a tetragonal form with the space group \emph{I}4/\emph{mcm}. In the $t$-MgO$_2$ phase (Fig. 2b), Mg is 8-coordinate. Here we see the same trend of change from 6-fold to 8-fold coordination as in the predicted B1-B2 transition in MgO. but in MgO$_2$ it happens at mere 53 GPa, compared to 490 GPa for MgO. Most remarkably, above 116 GPa the $t$-MgO$_2$ structure has a negative enthalpy of formation from MgO and O$_2$, indicating that $t$-MgO$_2$ becomes thermodynamically stable. Furthermore, its stability is greatly enhanced by pressure and its enthalpy of formation becomes impressively negative, -0.43 eV/atom, at 500 GPa ! We also examined the effect of temperature on its stability by performing quasi-harmonic free energy calculations. Inclusion of thermal effects clearly does not bring any qualitative changes in the formation free energy ($\Delta$$G$), and MgO$_2$ remains stable at high temperatures.
\begin{figure}
\begin{center}
\epsfig{file=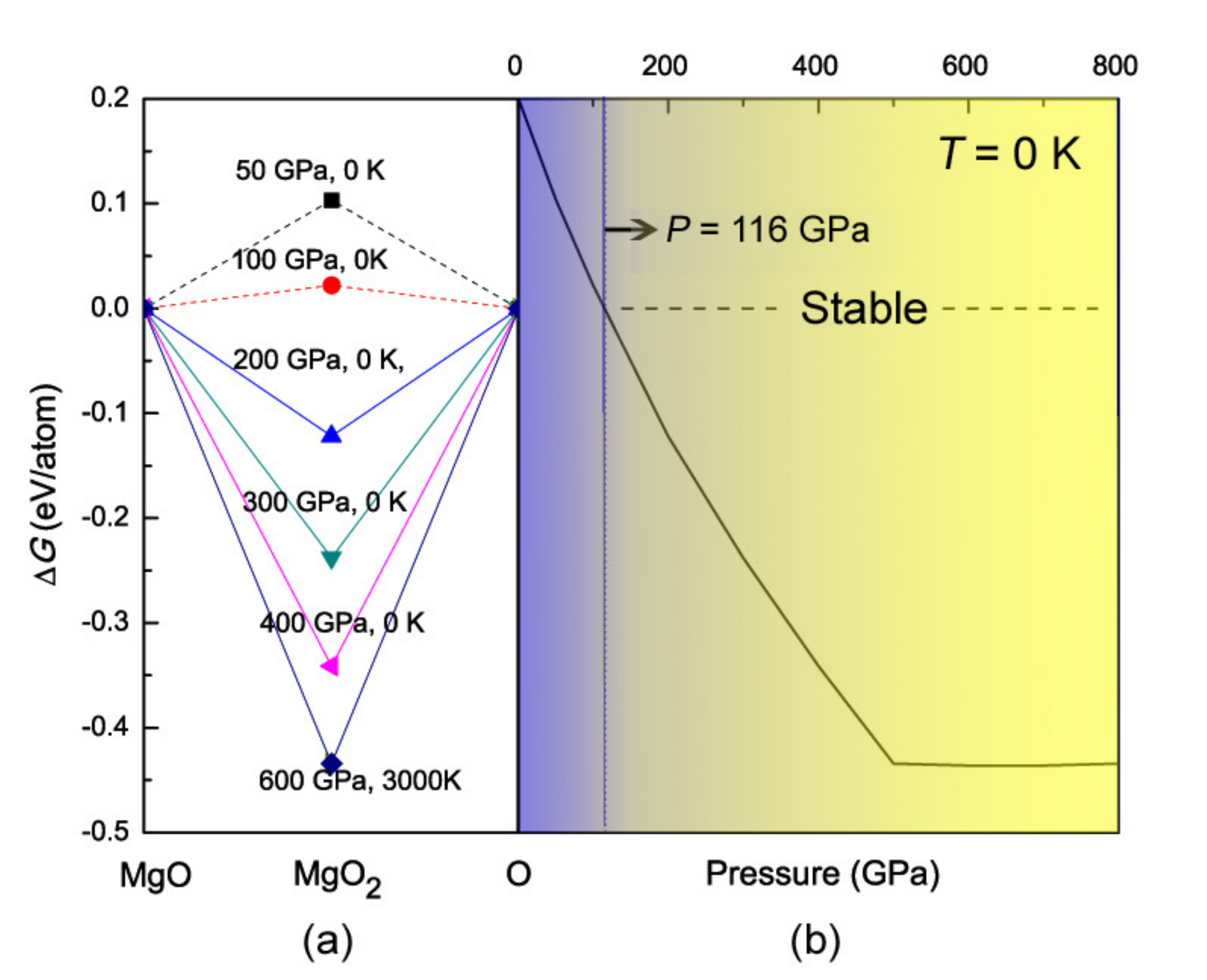,width=0.4\textwidth}
\caption{(a) Convex hull for the MgO-O system at high pressures; (b) the formation enthalpy as a function of pressure. For oxygen, the structures of the $\zeta$-phase \cite{11} and $\epsilon$-phase \cite{12} were used. For MgO, B1 and B2 phases were considered. $c$-MgO$_2$ was considered at 50 GPa, while $t$-MgO$_2$ was used at pressures higher than 50 GPa.}\label{fig1}
\end{center}
\end{figure}
\begin{figure}
\epsfig{file=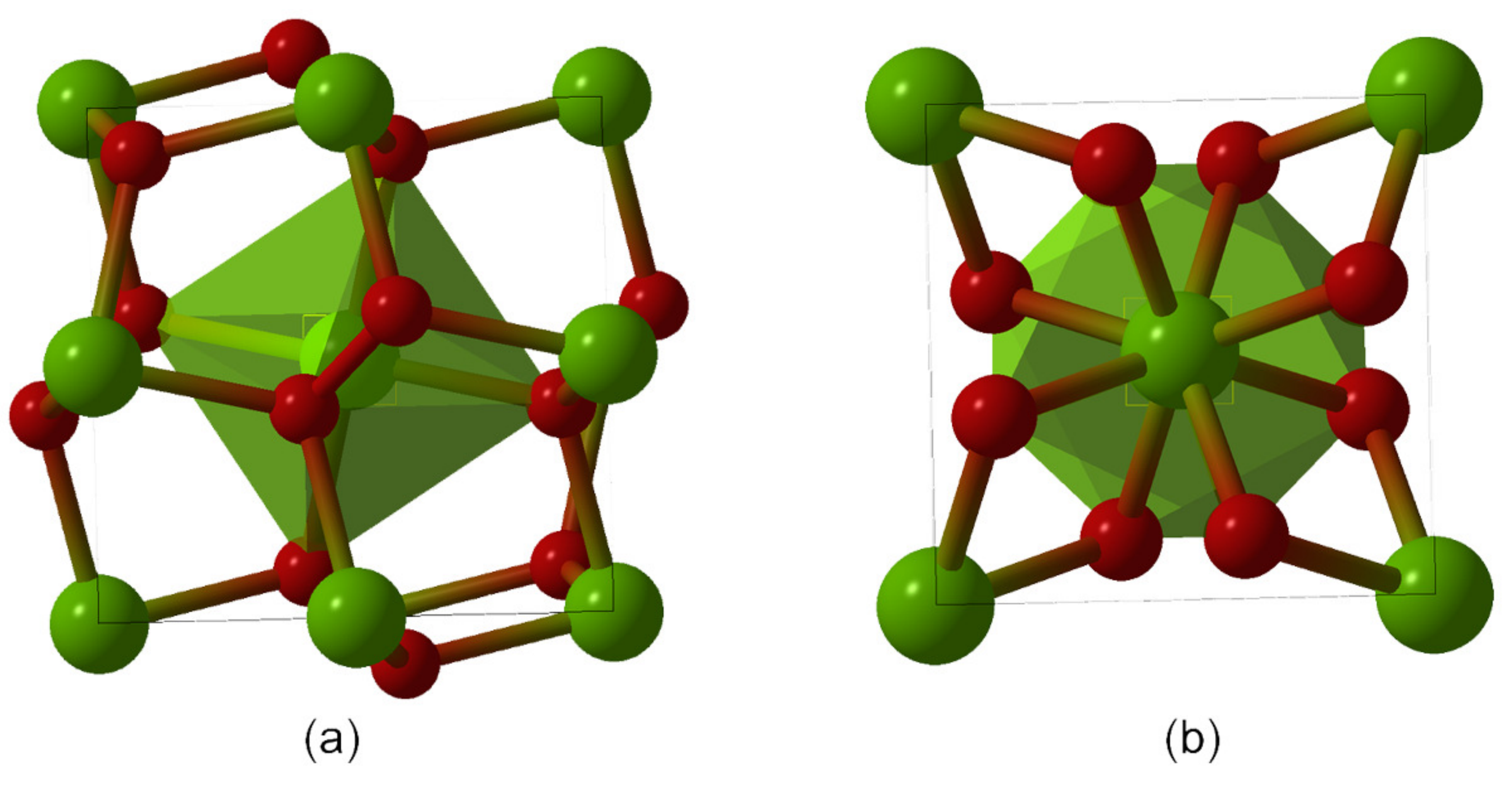,width=.4\textwidth}
\begin{center}
\caption{Crystal structures of (a) $c$-MgO$_2$ phase at 50 GPa, space group \emph{Pa}3, a=4.5236 \AA, Mg(0, 0, 0), O(0.4074, 0.4074, 0.4074); (b)$t$-MgO$_2$ at 500 GPa, space group \emph{I}4/\emph{mcm}, a=3.3773 \AA, c=3.9849 \AA, Mg(0, 0, 0.75), O(0.1260, 0.3740, 0.5). The green polyhedra are drawn to show the coordination environment for the selected Mg atoms (6-fold in $c$-MgO$_2$, and 8-fold in $t$-MgO$_2$).}\label{fig2}
\end{center}
\end{figure}

\subsection{Phase diagram of the Mg-MgO system: Mg$_3$O$_2$ is an exotic stable compound} For the Mg-rich part of Mg-O phase diagram, USPEX shows completely unexpected results. First of all, elemental Mg is predicted to undergo several phase transitions induced by pressure: hcp - bcc - fcc - sh. At ambient condition, Mg adopts the hcp structure, while bcc-Mg is stable from 50 GPa to 456 GPa, followed by the transition to fcc and simple hexagonal phase at 456 and 756 GPa, respectively. These results are in excellent agreement with previous studies \cite{13, 14, 15}. Mg-rich oxides, such as Mg$_2$O and Mg$_3$O$_2$ begin to show very competitive enthalpy of formation at pressures above 100 GPa. However, they are still not stable against decomposition into Mg and MgO, and their crystal structures could be thought as a combination of blocks of Mg and B1-MgO. This situation qualitatively changes at 500 GPa, where we find that Mg$_3$O$_2$ becomes thermodynamically stable. This new stable phase has a very unusual tetragonal structure with the space group \emph{P}4/\emph{mbm}.

This crystal structure can be viewed as a packing of O atoms and 1D-columns of almost perfect body-centered Mg-cubes. As shown in Fig. 3, there are two types of Mg atoms in the unit cell, Mg1 and Mg2 - Mg2 atoms form the cubes, filled by Mg1 atoms. Within the cubic columns, one can notice empty (Mg1)$_2$(Mg2)$_4$ clusters with the shape of flattened octahedra, with Mg-Mg distances ranging from 2.08 \AA\ (Mg1-Mg2) to 2.43 \AA\ (Mg2-Mg2). The coordination environments are quite different: each Mg1 is bonded to two Mg1 atoms and eight Mg2 atoms, while each Mg2 atoms is bonded to six O atoms (trigonal prismatic coordination). Oxygen atoms in \emph{t}-Mg$_3$O$_2$ are coordinated by eight Mg2 atoms.

What are the implications of these two compounds for planetary sciences? High pressures, required for their stability, are within the range corresponding to deep planetary interiors. In planetary interiors, reducing conditions dominate, related to the excess of metallic iron. However, given the diversity of planetary bodies it is not impossible to imagine that on some planets strongly oxidized environments can be present at depths corresponding to the pressure of 116 GPa (in the Earth this corresponds to depths below ~2600 km). At the more usual reducing conditions of planetary interiors, Mg$_3$O$_2$ could exist at pressures above 500 GPa in deep interiors of giant planets. There, it can coexist in equilibrium with Fe (but probably not with FeO, according to our DFT and DFT+U calculations of the reaction Fe + 3MgO $\rightarrow$ FeO + Mg$_3$O$_2$). According to our calculations (Fig. 4), Mg$_3$O$_2$ can only be stable at temperatures below 1800 K, which is too cold for deep interiors of giant planets; however, impurities and entropy effects stemming from defects and disorder might extend its stability field into planetary temperatures. Exotic compounds MgO$_2$ and Mg$_3$O$_2$, in addition to their general chemical interest, might be important planet-forming minerals in deep interiors of some planets.
\begin{figure}
\epsfig{file=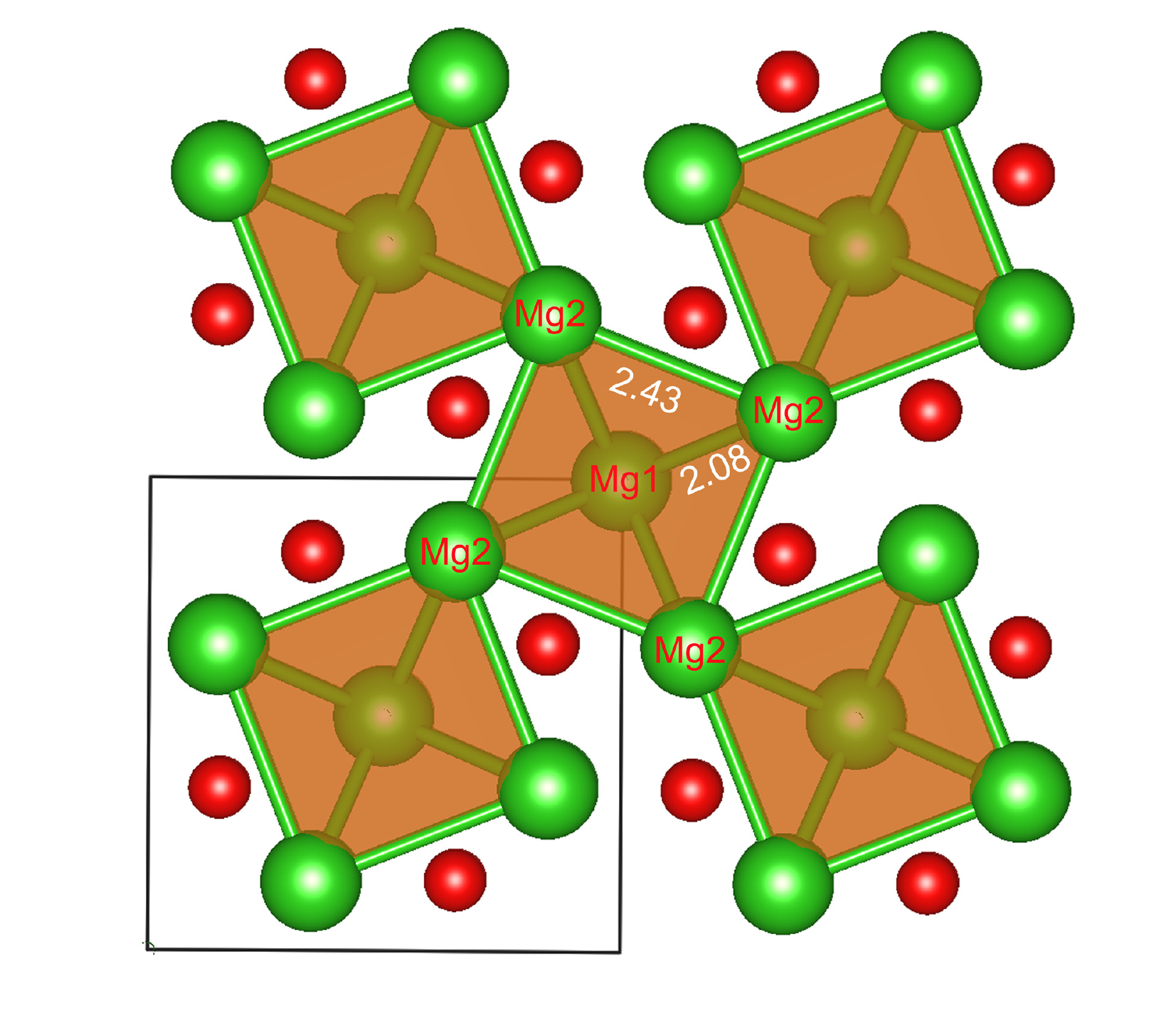,width=.25\textwidth}
\begin{center}
\caption{Crystal structures of $t$-Mg$_3$O$_2$ at 500 GPa, space group \emph{P}4/\emph{mbm}, a=4.5079 \AA, c=2.3668 \AA, Mg1(0.3494, 0.1506, 0.5); Mg2(0, 0, 0); O(0.8468, 0.6532, 0). This representation highlights the Mg octahedra.}
\label{fig3}
\end{center}
\end{figure}

\begin{figure}
\epsfig{file=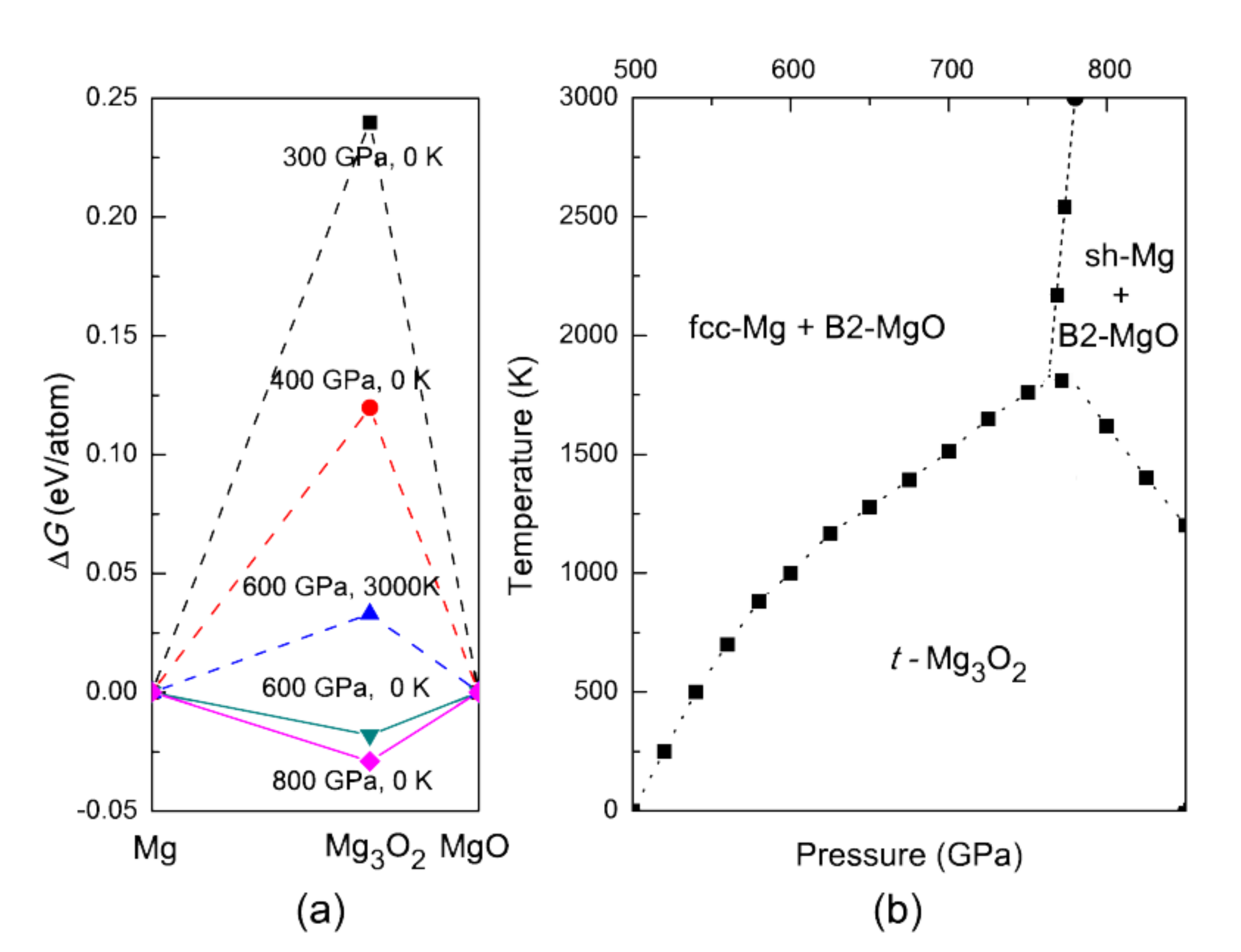,width=.5\textwidth}
\begin{center}
\caption{(a) Convex hull for the Mg-MgO system at high pressures. For Mg, the bcc, fcc and sh structures were used at their stability range. For MgO, B1 and B2 phases were considered. $t$-Mg$_3$O$_2$ structure was used in the whole investigated pressure range. (b) Predicted ${P-T}$ phase diagram for the Mg-MgO system.}
\label{fig4}
\end{center}
\end{figure}

\section{Electronic Structure}

Phonon calculations for Mg$_3$O$_2$ and MgO$_2$ at pressures of their stability show that no imaginary phonon frequencies exist throughout Brillouin zone, suggesting that these structures are dynamically stable. Together with our calculated thermodynamic functions, this suggests thermodynamic stability of these compounds. How are these exotic magnesium oxides stabilized at high pressure? To answer this question, we analyzed the electronic structure and chemical bonding for these compounds.

\begin{figure*}
\begin{center}
\epsfig{file=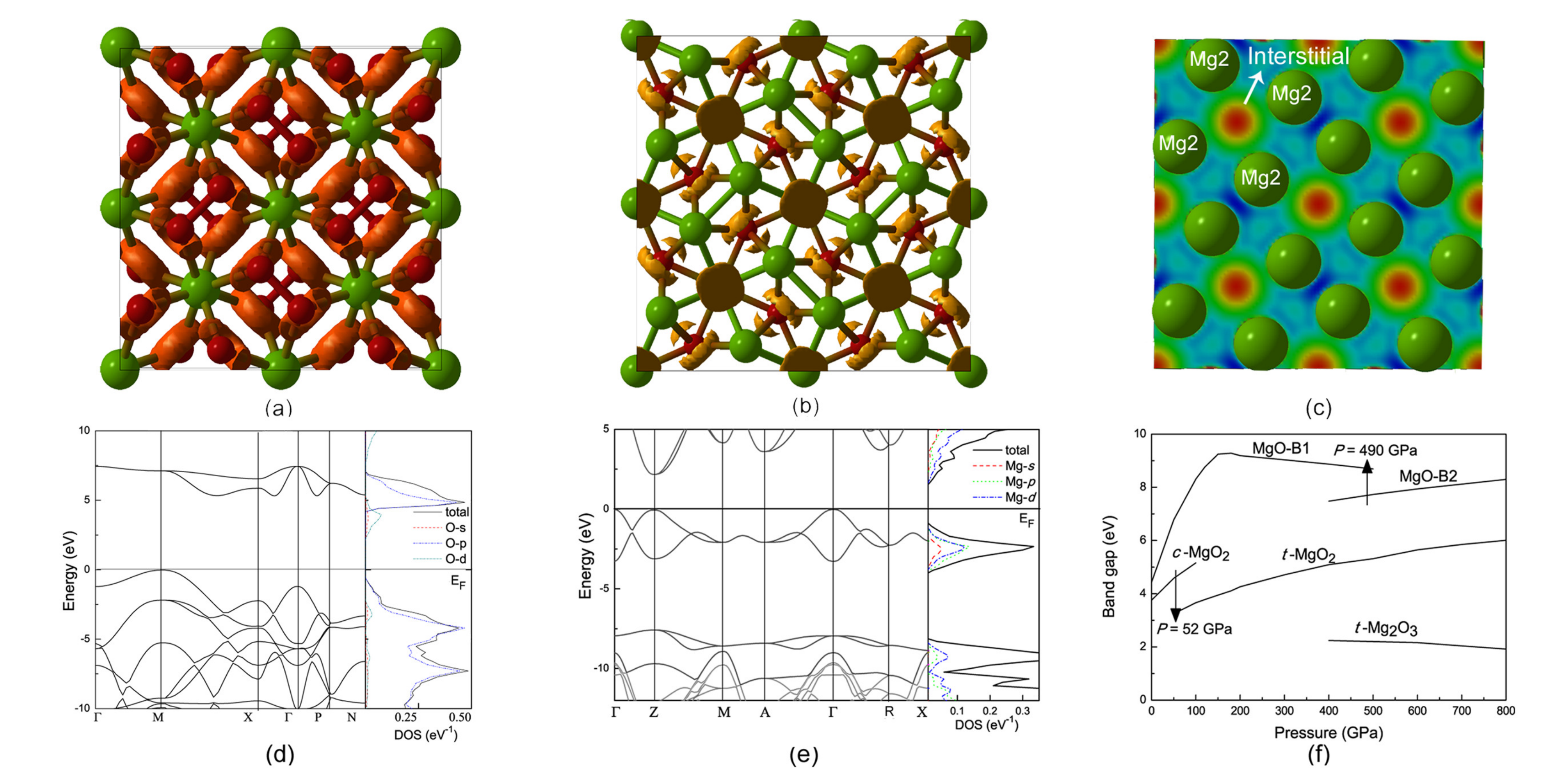,width=.95\textwidth}
\caption{ELF isosurface of a) $t$-MgO$_2$ structure; b) $t$-Mg$_3$O$_2$ structure at 500 GPa and $ELF$ = 0.83. c) Charge density distribution of $t$-Mg$_3$O$_2$ along c axis. Band structure and partial densities of states for d) $t$-tetragonal MgO$_2$ and e) $t$-Mg$_3$O$_2$. f) Band gaps as a function of pressures for various Mg-O compounds.}\label{fig5}
\end{center}
\end{figure*}

Electron Localization Function (ELF) gives information about the bonding character and valence electron configuration of atoms in a compound \cite{16}. The ELF pictures show asymmetry of the Mg-O bonds, indicating significantly ionic character of bonding. As shown in Fig. \ref{fig5}a, valence electrons in $t$-MgO$_2$ are mainly concentrated around O atoms. Charge transfer was also investigated on the basis of the electron density using Bader analysis \cite{17, 18}. In $t$-MgO$_2$, the net charge on Mg is +1.747 $e$, indicating the nearly complete transfer of Mg 3$s$ electrons to O atoms (just like Mg in MgO: Bader charges are +1.737 $e$ at 0 GPa and +1.675 $e$ at 600 GPa, respectively). Each O has almost 7 valence electrons (6.873 $e$), thus with the formation of singly bonded O-O dumbbell the octet rule is fulfilled. Each O-O dumbbell can be viewed as a peroxide-ion [O-O]$^{2-}$ with a closed-shell electronic configuration. 

The ELF distribution in $t$-Mg$_3$O$_2$ (Fig. \ref{fig5}b) also confirms strong charge transfer from Mg to O. However, we surprisingly found a very strong interstitial ELF maximum (\emph{ELF}=0.97) located in the center of Mg-octahedron (Fig. \ref{fig5}c). To obtain more insight, we performed Bader analysis. The resulting charges are +1.592 $e$ for Mg1, +1.687 $e$ for Mg2, -1.817 $e$ for O, and -1.311 $e$ for the interstitial electron density maximum. Such a strong interstitial electronic accumulation requires an explanation. Its electronic structure (Fig. \ref{fig5}e) exhibits intriguingly high occupancy of Mg-$p,d$-orbitals with overlapping energy ranges, which implies a strong Mg 3$p$-3$d$ hybridization. At high pressure, strong interstitial electron localization was found in alkali and alkaline-earth elements, for instance, sodium becomes insulator due to strong core-core orbital overlap \cite{19}. As a measure of size of the core region we use the Mg$^{2+}$ ionic radius (0.72 \AA \cite{20}), while the size of the valence electronic cloud is represented by the 3s orbital radius (1.28 \AA \cite{21}). In Mg$_3$O$_2$, Mg-Mg contacts at 500 GPa (2.08 A for Mg1-Mg2, 2.37 A for Mg1-Mg1 and 2.43 A for Mg2-Mg2) are only slightly shorter than the sum of valence orbital radii. Thus, the main reason for strong interstitial electronic localization is the formation of strong multicenter covalent bonds between Mg atoms; the core-valence expulsion (which begins at distances slightly longer than the sum of core and valence radii and increases as the distance decreases) could also be an important factor for valence electron localization. Strong Mg-Mg covalent bonding is not normally observed; the valence shell of the Mg atom only has a filled 3$s$$^2$ orbital, a configuration unfavorable for strong bonding. Under pressure, the electronic structure of the Mg atom changes ($p$- and $d$-levels become significantly populated), and strong covalent bonding can appear as a result of the $p$-$d$ hybridization. There is another way to describe chemical bonding in this unusual compound. We remind that Mg$_3$O$_2$ is anion-deficient compared with MgO; the extra localized electrons in Mg octahedron interstitial play the role of anions, screening Mg atoms from each other. These two descriptions are complementary.

\section{Conclusions}
In summary, we performed a systematic search for possible stoichiometries in the Mg-O system at pressures up to 850 GPa. Other than the well-known compound, MgO, we found that two more stoichiometries (MgO$_2$ and Mg$_3$O$_2$) become thermodynamically stable at pressures above 116 GPa and 500 GPa, respectively. Our analysis reveals that bonding in both $t$-MgO$_2$ and $t$-Mg$_3$O$_2$ exhibits significantly ionic character. MgO$_2$ is stabilized by the formation of the peroxide [O-O]$^{2-}$ anion, while the strong electron localization in Mg octahedron plays the role of an additional anion and makes the anion-deficient compound Mg$_3$O$_2$ stable. Compared with MgO, which is a wide gap insulator (DFT band gap 7.73 eV at 500 GPa - see Fig. \ref{fig5}f), Mg$_3$O$_2$ is predicted to have a much narrower band gap (it is a semiconductor with a DFT gap of 2.20 eV), while MgO$_2$ is an insulator with the DFT band gap of 5.35 eV; note that DFT gaps are always lower bound to the true gaps. These two compounds might exist in interiors of other planets, and are thus potentially important for our fundamental understanding of the universe - in addition to presenting striking new chemical phenomena.

\section{Methods: Crystal Structure Search} Searches for the stable compounds and structures were performed using evolutionary algorithm USPEX, as implemented in the USPEX code\cite{22, 23}. This was done in conjunction with \textit{ab initio} structure relaxations based on density functional theory (DFT) within the Perdew-Burke-Ernzerhof (PBE) generalized gradient approximation (GGA)\cite{24}, as implemented in the VASP code\cite{25}. For structural relaxation, we used the all-electron projector-augmented wave (PAW) method \cite{26} and plane wave basis set with the 600 eV kinetic energy cutoff; Brillouin zone was sampled by Monkhorst-Pack meshes with the resolution 2$\pi$ $\times$ 0.06 \AA$^{-1}$. Such calculations provide an excellent description of the known structures (Mg, O$_2$, MgO) and their energetics. The most significant feature of USPEX we used in this work is the capability of searching for a specific area of the composition space - as opposed to the more usual structure predictions at fixed chemical composition. The desired composition space is described via building blocks (for example, search for all compositions in a form of [\textit{x}Al$_2$O$_3$+\textit{y}MgO] or [\textit{x}Mg+\textit{y}O]). During the initialization USPEX samples the whole range of compositions of interest randomly and sparsely. Chemistry-preserving constraints in the variation operators are lifted and replaced by block-correction scheme which ensures that a child structure is within the desired area of composition space. A new ``chemical transmutation" operator is introduced \cite{27, 28}. Stable compositions are determined using the convex hull construction: a compound is thermodynamically stable if the enthalpy of its decomposition into any other compounds is positive. To ensure that the obtained structures are dynamically stable, we calculated phonon frequencies across the Brillouin zone using the finite-displacement approach as implemented in the Phonopy code\cite{29}. We also explored the effects of temperature using the quasiharmonic approximation; for each structure, phonons were calculated at 10 - 15 different volumes.

\begin{acknowledgments}
Calculations were performed at the supercomputer of Center for Functional Nanomaterials, Brookhaven National Laboratory, which is supported by the U.S. Department of Energy, Office of Basic Energy Sciences, under contract No. DE-AC02-98CH10086, at the Joint Supercomputer Center of the Russian Academy of Sciences and on Skif supercomputer (Moscow State University). This work is funded by DARPA (grants No. N660011014037, W31P4Q1210008), National Science Foundation (No. EAR-1114313).
\end{acknowledgments}

\end{document}